\documentclass[%
 preprint,
 superscriptaddress,
preprintnumbers,
nofootinbib,
 amsmath,amssymb,
11pt
]{revtex4-2}
\pdfoutput=1
\usepackage{amsmath}
\usepackage{amssymb}
\usepackage{slashed}
\usepackage{graphicx}
\usepackage{graphics}
\usepackage{epsfig}
\usepackage{color}
\usepackage{xcolor}
\usepackage{soul}
\usepackage{hyperref}
\usepackage[section]{placeins}
\usepackage{mathrsfs}
\usepackage{dsfont}
\usepackage{comment}

\newcommand{\be}{\begin{equation}}
\newcommand{\ee}{\end{equation}}
\newcommand{\bea}{\begin{eqnarray}}
\newcommand{\eea}{\end{eqnarray}}

\begin{document}
\title{
Parameterizations of black-hole spacetimes beyond circularity
}

\author{H\'elo\"ise Delaporte}
   \email{hdel@sdu.dk}
   \affiliation{CP3-Origins, University of Southern Denmark, Campusvej 55, DK-5230 Odense M, Denmark}

\author{Astrid Eichhorn}
   \email{eichhorn@sdu.dk}
   \affiliation{CP3-Origins, University of Southern Denmark, Campusvej 55, DK-5230 Odense M, Denmark}

\author{Aaron Held}
  \email{aaron.held@uni-jena.de}
\affiliation{Theoretisch-Physikalisches Institut, Friedrich-Schiller-Universit\"at Jena, Max-Wien-Platz 1, 07743 Jena, Germany}
\affiliation{The Princeton Gravity Initiative, Jadwin Hall, Princeton University, Princeton, New Jersey 08544, U.S.}

\begin{abstract} 
We discuss parameterizations of black-hole spacetimes in and beyond General Relativity in view of their symmetry constraints: within the class of axisymmetric, stationary spacetimes, we propose a parameterization that includes non-circular spacetimes, both in Boyer-Lindquist as well as in horizon-penetrating coordinates. We show how existing parameterizations, which make additional symmetry assumptions (first, circularity; second, a hidden constant of motion), are included in the new parameterization.
Further, we explain why horizon-penetrating coordinates may be more suitable to parameterize non-circular deviations from the Kerr geometry.

Our investigation is motivated by our result that the regular, spinning black-hole spacetimes proposed in \cite{Eichhorn:2021etc,Eichhorn:2021iwq} are non-circular. 
This particular deviation from circularity can result
in cusps, a dent and an asymmetry in the photon rings surrounding the black-hole shadow.

Finally, we explore a new class of non-circular deviations from Kerr black holes, which promote the spin parameter to a function, and find indications that regularity cannot be achieved in this setting. This result strengthens the case for regular black holes based on a promotion of the mass parameter to a function.
\end{abstract}

\maketitle

\section{Introduction}
From a fundamental point of view, it is not a question \emph{whether} the Kerr paradigm breaks down for  astrophysical black holes. It is only a question, \emph{when} (and how) it breaks down.
Current and future observations of the near-horizon geometry of black-hole spacetimes \cite{paper1, paper6,EventHorizonTelescope:2021bee} offer an exciting possibility to i) test the validity of General Relativity (GR) for an unprecedented range of scales, ii) learn about the true nature of black holes and iii) potentially even catch a glimpse of quantum gravitational effects; see \cite{EventHorizonTelescope:2021dqv} for first observational constraints on black-hole ``charges" and, e.g., \cite{Giddings:2016btb,Giddings:2019jwy,Held:2019xde,Broderick:2021ohx,Bacchini:2021fig} for further studies.

In order to achieve these goals, it is crucial to understand how black holes beyond GR ``look like", i.e., what features their shadows have. This can be done in specific theories beyond GR, but such an approach is not comprehensive. 
In contrast, parameterized approaches of deviations  from the Kerr metric can provide a comprehensive catalogue of black-hole shadows beyond GR. To achieve comprehensiveness, the parameterizations have to be as general as necessary to cover all relevant cases. 

In this paper, we review that commonly used parameterizations, such as \cite{1979GReGr..10...79B,Johannsen:2011dh,Johannsen:2013szh,Cardoso:2014rha,Konoplya:2016jvv}, make an important assumption about the structure of black-hole spacetimes, namely circularity. Circularity is an isometry that links time-reflections to reflections in the azimuthal angle. It has a number of consequences for black-hole spacetimes (and consequently their shadows) and does not hold in all settings beyond GR.

Thus, we put forward a more general parameterization. 
We first argue why Boyer-Lindquist coordinates are not well-suited to describe non-circular black holes beyond GR. The reason is that these parameterizations rely on delicate cancellations to avoid curvature singularities at the horizon \cite{Johannsen:2013szh, Johannsen:2013rqa, Cardoso:2014rha, Held:2021vwd}. Thus, their use for  black holes is complicated by additional (in general differential) constraints on deviations from the Kerr metric. 
We therefore propose a new parameterization of black holes in horizon-penetrating coordinates, more specifically,   in ingoing Kerr coordinates. This horizon-penetrating parameterization contains  circular spacetimes as special subclasses, but is more general and allows for deviations from circularity. We find that circularity is, however, difficult to impose in horizon-penetrating coordinates, which motivates  the use of Boyer-Lindquist coordinates for circular spacetimes.

Our study is motivated by the new result that a specific deviation from circularity is linked to a locality principle \cite{Eichhorn:2021iwq} and in turn results in particular image features in black-hole shadows~\cite{Eichhorn:2021etc}. Further, deviations from circularity have recently been constrained theoretically \cite{Xie:2021bur}, where it has been shown that under certain assumptions,  theories which are based on perturbative deviations from GR  contain only circular spacetimes as solutions. Taken together, these results suggest that there may be a (not necessarily one-to-one) link from specific deviations from circularity to certain image features (those found in \cite{Eichhorn:2021etc,Eichhorn:2021iwq}) and to certain classes of theories. Naturally, such a link relies on several assumptions and we cannot exclude that other spacetime-properties can result in similar observational features. Nevertheless, this link is an important motivation to explore non-circular parameterizations and we propose that future next-generation Event Horizon Telescope (EHT) tests of GR, which may benefit from increased image resolution, take non-circular deviations into account.

This paper is structured as follows: in Sec.~\ref{sec:non-circularity}, we review the property of circularity and highlight examples of non-circular black-hole spacetimes. In Sec.~\ref{sec:noncircimage}, we establish that specific  deviations from circularity result in image features in shadow images of a family of regular black-hole spacetimes. This family of black-hole spacetimes was introduced in \cite{Eichhorn:2021etc,Eichhorn:2021iwq}, based on a locality principle founded in fundamental-physics considerations. Here, we show that these locality-based spacetimes are not circular, which establishes a tentative link between the locality principle, non-circularity, as well as cusps, a dent and asymmetry in the photon rings surrounding the black-hole shadow.
These phenomenological considerations motivate us to review the construction of parameterizations of black-hole spacetimes. In Sec.~\ref{sec:circular}, we review how circular parameterizations are constructed, both with and without a hidden constant of motion. In Sec.~\ref{sec:hp} and Sec.~\ref{sec:BLgen}, we introduce new parameterizations which include non-circular spacetimes. We first provide a parameterization in horizon-penetrating coordinates, and then a parameterization in coordinates which reduce to Boyer-Lindquist coordinates in the Kerr limit. In both sets of coordinates, we write the black-hole spacetime in terms of deviations from Kerr spacetime. In horizon-penetrating coordinates, we also explore a specific subset of spacetimes, namely those which are regularized by a particular form of ``quantum hair" that upgrades parameters of the spacetime (mass and spin) to functions. Interestingly, we find indications that spin ``hair" on its own cannot regularize a black-hole spacetime. This singles out regular black holes based on a modification of the mass-function, as in \cite{Bardeen:1968,Dymnikova:1992ux,Bonanno:1998ye,Hayward:2005gi,Nicolini:2008aj,Reuter:2010xb,Bambi:2013ufa,Azreg-Ainou:2014pra,Haggard:2014rza,Toshmatov:2014nya,Ghosh:2014hea,Abdujabbarov:2016hnw,Torres:2017gix,Platania:2019kyx,Simpson:2019mud,Ashtekar:2018lag,Kumar:2019ohr,Nicolini:2019irw,Shaikh:2019fpu,Contreras:2019cmf,Liu:2020ola,Lima:2021las,Junior:2021atr,Mazza:2021rgq,Eichhorn:2021etc,Eichhorn:2021iwq}.
Finally, we conclude and list open questions in Sec.~\ref{sec:conclusions}.

\section{Non-circular spacetimes: fundamentals and examples}\label{sec:non-circularity}
Restricting to the observed four dimensions, the most general metric has ten independent components that can each depend on four coordinates. For axisymmetric, stationary and asymptotically flat spacetimes, one expects that the form of the metric components simplifies. In addition, one might require stronger conditions that reduce the number of metric components, namely i) circularity and ii) an additional (Carter-like) hidden constant of motion, both of which we explain below. We will focus on black holes but also want to account for black-hole mimickers, i.e., we are also interested in spacetimes that do not feature a horizon. Thus, we do not impose the existence of a horizon as an additional constraint on the metric.

Axisymmetric and stationary spacetimes have two Killing vectors $\xi_1$ and $\xi_2$, which commute \cite{Carter:1970ea}. Therefore, there are adapted coordinates  in which the metric only depends on the two non-Killing coordinates.

 If circularity holds (as it does in vacuum GR), then it restricts the geometry further. 
Circularity holds  \cite{Papapetrou:1966zz} if and only if
\bea
\xi_1^{[\mu} \xi_2^{\nu}\nabla^{\kappa}\xi_1^{\lambda]} &=& 0 \mbox{ at at least one point,}\\
\xi_2^{[\mu} \xi_1^{\nu}\nabla^{\kappa}\xi_2^{\lambda]} &=& 0 \mbox{ at at least one point,}\\
\xi_1^{\mu} R_{\mu}^{\,\, [\nu}\xi_2^{\kappa}\xi_1^{\lambda]} &=& 0 \mbox{ everywhere,}\label{eq:circular1}\\
\xi_2^{\mu} R_{\mu}^{\,\, [\nu}\xi_1^{\kappa}\xi_2^{\lambda]} &=& 0 \mbox{ everywhere} \label{eq:circular2}.
\eea
Herein, $\nabla$ denotes the covariant derivative, $R_{\mu\nu}$ the Ricci tensor and square brackets denote antisymmetrization of all enclosed indices.
Since we focus on asymptotically flat spacetimes, axisymmetry implies the existence of an axis of rotation on which the Killing vector, say $\xi_2$, associated to  azimuthal rotations, must vanish. Hence, for the spacetimes in our paper, the two first conditions always hold.
Together with the latter two conditions, they imply the existence of an isometry of the spacetime. In Boyer-Lindquist coordinates, this isometry simplifies to the simultaneous transformation of $t \rightarrow -t$ and $\phi_\text{BL} \rightarrow -\phi_\text{BL}$, see, e.g., \cite{Ayon-Beato:2005mje}.

Circularity also appears in relation to the existence of closed photon orbits. Every stationary, axisymmetric, and asymptotically flat black-hole spacetime that is also circular must admit at least two planar closed photon orbits -- one with and one against the rotation of the black hole \cite{Cunha:2020azh}. It is not known whether the proof can be generalized beyond circularity.

For axisymmetric and stationary black holes, circularity implies that the angular velocity is constant on the event horizon  \cite{Frolov:1998wf}. For GR, this property is encoded in Hawking's rigidity theorem \cite{Hawking:1973uf}. Non-circular black holes can instead have event horizons on which the angular velocity is not constant. 

In vacuum GR, circularity holds since $R_{\mu\nu}=0$. Beyond vacuum GR, circularity need not hold.
Viewed from the physics perspective, black holes as rotating compact objects need not have a uniformly rotating event horizon.
Viewed from the symmetries perspective, black-hole spacetimes with stationarity and axisymmetry need not satisfy additional isometries\footnote{From an astrophysical perspective, where black holes form from collapse, or from a theoretical perspective accounting for Hawking radiation, even stationarity appears as a too strong symmetry assumption, at least on long time scales compared to the gravitational time-scale associated to a black hole of astrophysical mass.}.
\\

Some metrics beyond GR satisfy circularity \cite{Nakashi:2020phm,Xie:2021bur}, but even within GR, non-circular spacetimes exist.
For instance, toroidal magnetic fields and convective motion are both known to result in non-circular metrics for neutron stars~\cite{Ioka:2003dd, Ioka:2003nh, Birkl:2010hc}. Further, \cite{Vera:2003cn} suggests that non-circular interior solutions for compact rotating bodies can be matched onto circular external solutions in GR. Beyond GR, non-circular spinning black holes
were recently derived as solutions to modified gravitational dynamics: \cite{Minamitsuji:2020jvf,Anson:2020trg,BenAchour:2020fgy} construct solutions in particular scalar-tensor or vector-tensor theories.

A different route has been followed in \cite{Held:2019xde}, which constructs a black-hole spacetime inspired by asymptotically safe gravity. While the explicit construction in \cite{Held:2019xde} contains a curvature singularity at the horizon, due to the use of Boyer-Lindquist coordinates (see \cite{Held:2021vwd}), the construction principle can be implemented in horizon-penetrating coordinates, resulting in a non-circular spacetime. The construction in \cite{Eichhorn:2021iwq} is an explicit example.

Further, within a parameterized approach, some non-circular spacetimes have been included in the framework of the ``bumpy Kerr metrics" in \cite[Eq.~(2.36)]{Vigeland:2009pr} and \cite[Eq.~(1.5)]{Vigeland:2010xe}.

Finally, a principled-parameterized approach has been developed in \cite{Eichhorn:2021etc,Eichhorn:2021iwq}, where deviations from Kerr black holes were restricted by a locality principle. The resulting regular, spinning black holes are non-circular and in \cite{Held:2021vwd} it has been argued that alternative regularizations, which are circular, violate the locality principle. 

\section{Deviations from circularity and shadow images}
\label{sec:noncircimage}

\subsection{Non-circularity and the principled-parameterized approach to black-hole spacetimes}

We now review the principled-parameterized approach to black-hole spacetimes \cite{Eichhorn:2021etc,Eichhorn:2021iwq}, in which a set of fundamental principles (see below) results in regular spacetimes with specific properties. Here, we show that these spacetimes are non-circular. We also discuss how the deviation from circularity imprints itself on image features in the black-hole shadow.

The principled-parameterized approach to black-hole spacetimes combines parameterizations of the metric with principles inspired by fundamental physics. In \cite{Eichhorn:2021etc,Eichhorn:2021iwq}, the two key principles\footnote{An additional simplicity principle is spelled out in detail in \cite{Johannsen2022}.} were regularity, i.e., the absence of curvature singularities, and locality, i.e., a connection to local curvature scales. The locality principle is motivated by an effective-field-theory approach to new physics in gravity: it states that where the local spacetime curvature is small, GR is a good approximation, such that the spacetime of a spinning black hole is locally equivalent to the Kerr spacetime in low-curvature regions. Conversely, where the local spacetime curvature is large, GR becomes an increasingly bad approximation, such that the spacetime deviates from the Kerr spacetime significantly. To implement this principle, all deviations from the Kerr spacetime are written as functions of
\be
K_{\rm GR}(r, \chi) = \frac{48 M^2}{\left(r^2+a^2\chi^2 \right)^3},
\label{eq:localcurvature}
\ee
where $r$ is the radial coordinate and $\chi=\cos \theta$ with $\theta$ the polar angle (in, for instance, ingoing Kerr coordinates).
$K_{\rm GR}$ forms an enveloping function for the absolute values of the independent Riemann curvature invariants in Kerr spacetime \cite{Eichhorn:2021iwq}. To combine the locality principle and the regularity principle,  it is enough to upgrade the mass parameter to a mass function $M(K_{\rm GR}(r,\chi))$ which depends on $K_{\rm GR}$ and goes to zero sufficiently fast in high-curvature regions, where $K_{\rm GR} \rightarrow \infty$.

The line element in the principled-parameterized approach that implements  locality and regularity is then given by
\bea
ds_{\rm reg,\, local}&=&-\frac{r^2-2M(r, \chi) r +a^2 \chi^2}{r^2+a^2 \chi^2}du^2 +2\,du\, dr - 4\frac{M(r,\chi) a r}{r^2+a^2\chi^2}\left(1-\chi^2 \right) du\, d\phi\nonumber\\
&{}&- 2a\left(1-\chi^2 \right)dr\, d\phi + \frac{r^2+a^2\chi^2}{1-\chi^2}d\chi^2 \nonumber\\
&{}&+ \frac{1-\chi^2}{r^2+a^2\chi^2}\left(\left(a^2+r^2\right)^2 - a^2\left(r^2-2M(r, \chi)r+a^2 \right)\cdot \left(1-\chi^2 \right) \right)d\phi^2,\label{eq:dsregloc}
\eea
in ingoing Kerr coordinates $u$, $r$, $\chi=\cos \theta$, $\phi$. As an example, one may choose 
$M_{\rm exp}(K_{\rm GR})=M\,e^{-(\ell_\textrm{NP}^4K_\textrm{GR})^{1/6}}$,
where $\ell_{\rm NP}$ is the length scale of new physics, and $M$ without arguments refers to the mass parameter.

Because the local curvature, $K_{\rm GR}$, is largest in the equatorial plane (at any given value of $r$), deviations from the Kerr spacetime are largest in the equatorial plane. The deviations correspond to an increase in compactness, which occurs, because regularity implies a decrease of gravitational strength. Thus, the location of the event horizon cannot be described just by $r_H= \rm const$, but needs to be $\chi$-dependent, i.e., $r_H=r_H(\chi)$. In the equatorial plane, the event horizon features a dent, i.e., $r_H(\chi)$ is smallest at $\chi=0$. This dent is absent if one gives up the requirement of locality and just demands regularity, as, e.g, in \cite{Bambi:2013ufa,Azreg-Ainou:2014pra,Toshmatov:2014nya,Ghosh:2014hea,Abdujabbarov:2016hnw,Torres:2017gix,Kumar:2019ohr,Shaikh:2019fpu,Contreras:2019cmf,Liu:2020ola,Lima:2021las,Mazza:2021rgq}. These works construct regular spinning black holes via the Janis-Newman algorithm \cite{Newman:1965tw}, see also \cite{Gurses:1975vu,Drake:1998gf}, which necessarily results in a hidden constant of motion \cite{Azreg-Ainou:2014pra,Shaikh:2019fpu,Junior:2020lya} and thus in circularity, see Sec.~\ref{sec:circular} below.
\\

\subsection{Non-circularity and image features}
\label{sec:cuspsanddents}

\begin{figure}
	\begin{center}
		\includegraphics[width=0.48\linewidth]{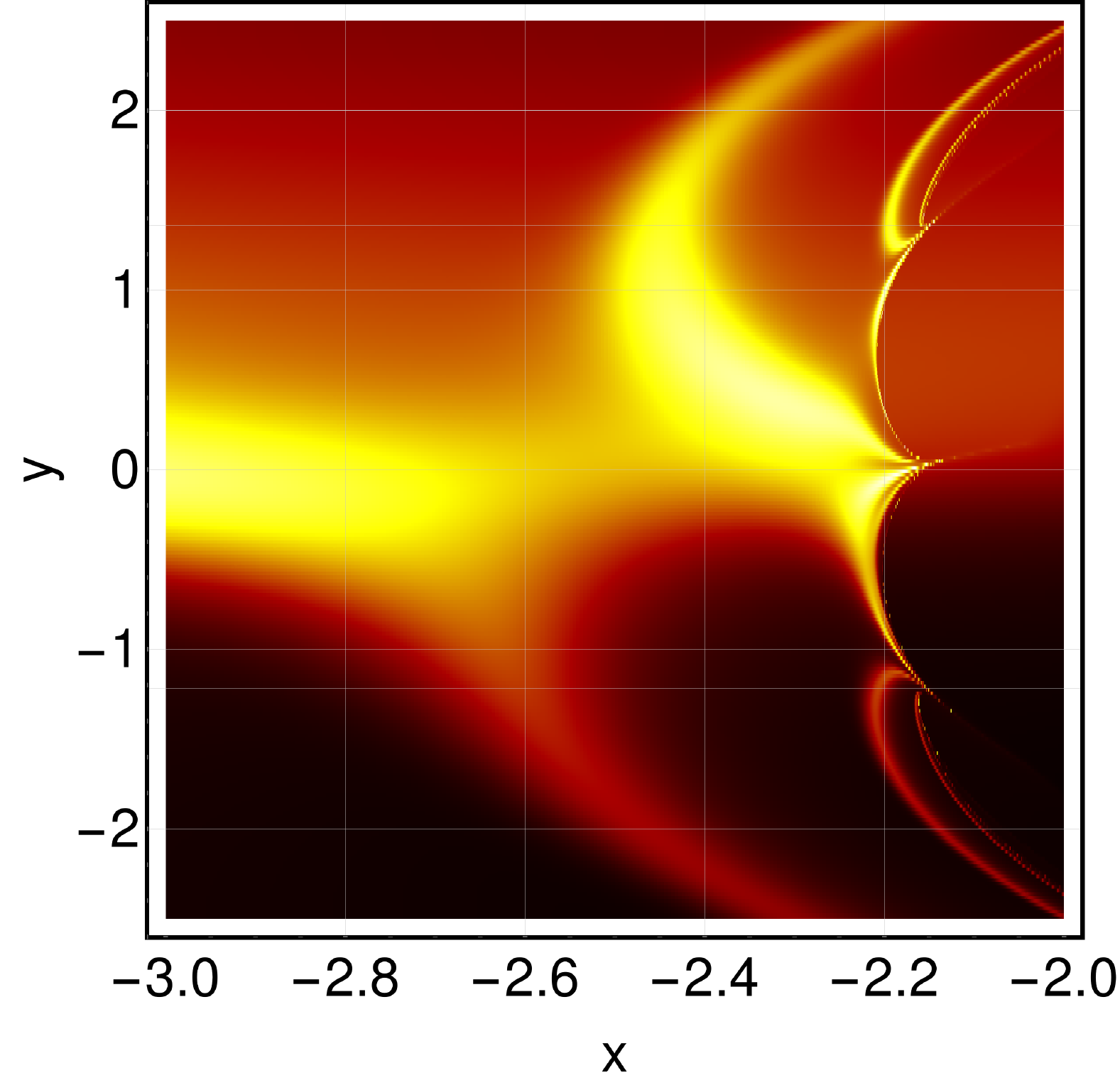}\hspace*{1em}
		\includegraphics[width=0.48\linewidth]{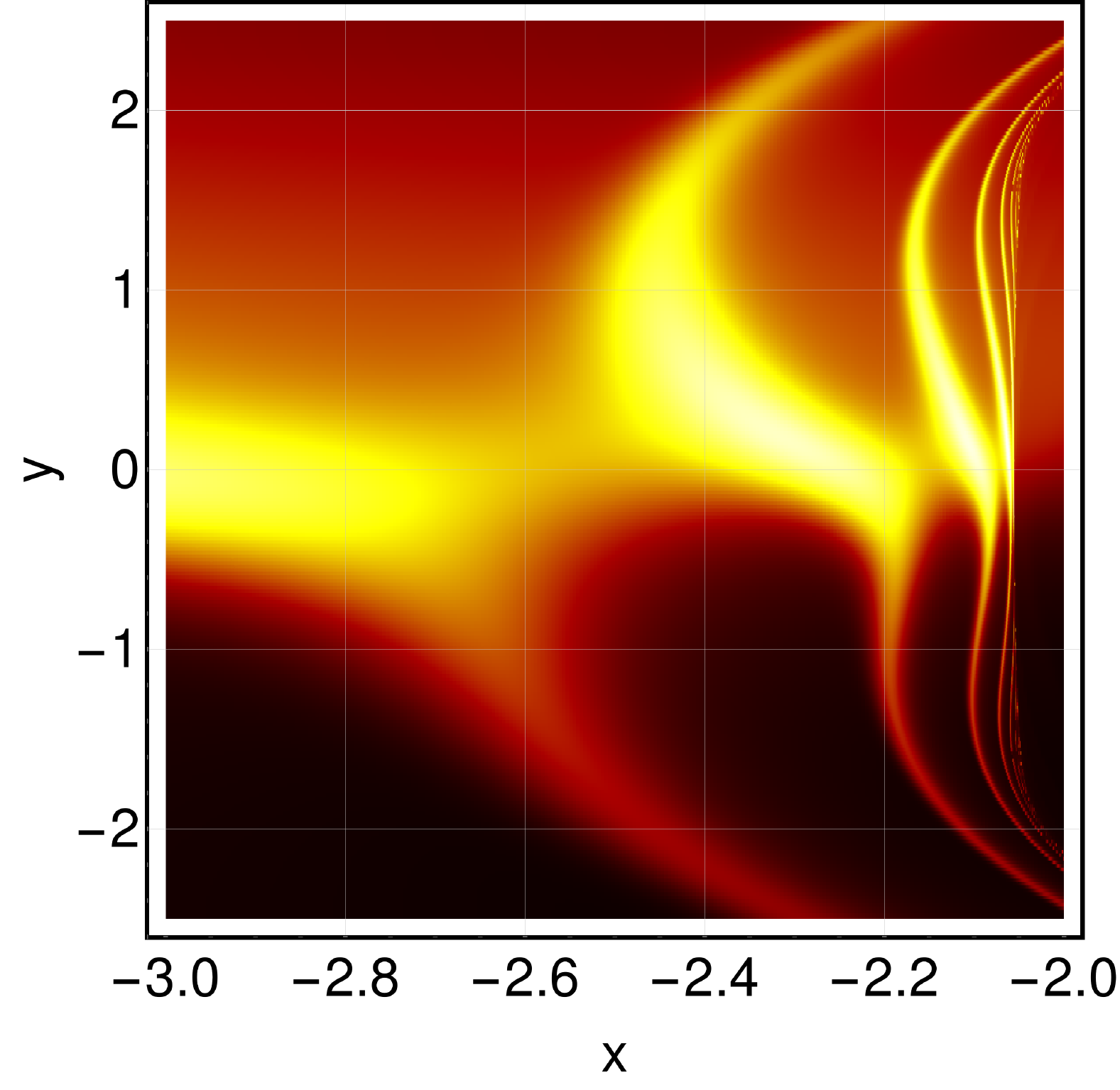}
	\end{center}
	\vspace*{-5pt}
	\caption{
	\label{fig:noncircular-vs-circular}
		Detailed view of the prograde image side (spacetime spinning towards the observer) for a non-circular (left panel) and a circular (right panel) regular black hole. Successive photon rings stack exponentially towards the shadow boundary from left to right in each image. The images are obtained by numerical ray tracing and radiative transfer of emission from a modelled accretion disk, cf.~\cite{Eichhorn:2021iwq} for details. The image intensity is normalized to the brightest image point. The non-circular and the circular spacetime are given in Eq.~\eqref{eq:dsregloc} with a mass function $M_{\rm non-circular}(K_{\rm GR})=M\,e^{-(\ell_\textrm{NP}^4K_\textrm{GR}(r,\chi))^{1/6}}$ and $M_{\rm circular}(K_{\rm GR})=M\,e^{-(\ell_\textrm{NP}^4K_\textrm{GR}(r,0))^{1/6}}$, respectively. We choose a large spin $a=0.9 M$ and a near-extremal new-physics scale $\ell_\text{NP}=0.1188 M$. All quantities are given in units of the classical (asymptotic) black-hole mass $M$.
	}
\end{figure}

Image features can be connected (though not necessarily in a one-to-one-way) to fundamental principles of spacetime. For instance, the images of black-hole spacetimes based on the regularity and the locality principle in \cite{Eichhorn:2021etc,Eichhorn:2021iwq} are characterized by:
\begin{enumerate} 
\item[(i)] cusps in the shadow boundary and the photon rings,
\item[(ii)] a dent \footnote{By a dent, we refer to the shadow boundary in the $y=0$ image axis lying  strictly inside its convex hull. In particular, the shadow boundary of extremal Kerr at edge-on inclination does not feature a dent.} in (the photon rings surrounding) the shadow boundary, 
\item[(iii)] broken reflection symmetry about the $y=0$ image axis at non-edge-on (and non-face-on) inclination, 
\end{enumerate}
cf.~left-hand panel in~Fig.~\ref{fig:noncircular-vs-circular}. All three features are absent for spacetimes that violate the locality principle, but are otherwise constructed as in \cite{Eichhorn:2021etc,Eichhorn:2021iwq}, see right-hand panel in~Fig.~\ref{fig:noncircular-vs-circular}. At the same time, these ``non-local" mass functions result in circular spacetimes. 

In short, the deviation from circularity in Eq.~\eqref{eq:dsregloc} results in specific image features. This does not hold for all deviations from circularity: one may easily construct deviations of the metric from the Kerr metric, which, though not necessarily motivated by fundamental physics, introduce a deviation from circularity that is localized far away from horizon and photonsphere and therefore does not lead to the same image features as the above example.

Further, the converse (i.e., absence of these image features in circular spacetimes) need not necessarily be true -- at least, we cannot provide a proof of this. However, we are only aware of circular examples that show one of the three image features, but not all three in combination.

First, many specific examples of circular black holes do not exhibit any of the three image features. For instance, the shadow of the metric \cite{Johannsen:2013szh}, which is circular and features a hidden constant of motion, is characterized only by an overall deformation of the prograde shadow boundary, but does not exhibit any of the three features \cite{Johannsen:2013vgc}. Similarly, the regular black holes in \cite{Bambi:2013ufa,Azreg-Ainou:2014pra,Toshmatov:2014nya,Ghosh:2014hea,Abdujabbarov:2016hnw,Torres:2017gix,Kumar:2019ohr,Shaikh:2019fpu,Contreras:2019cmf,Liu:2020ola,Junior:2021atr,Mazza:2021rgq} are circular and also characterized by an absence of these three image features and only show an overall deformation. 
In contrast, the circular spacetimes in \cite{Cunha:2017eoe,Wang:2017hjl} feature cusps in their shadow boundary (albeit of different type), but no dent in the $y=0$ image axis \cite{Cunha:2017eoe,Wang:2017hjl}. Conversely, the circular spacetimes in \cite{Khodadi:2021gbc}, based on the parameterization \cite{Johannsen:2011dh}, can feature a dent in the shadow boundary, but no cusps.

Thus, searching for such image features with EHT or ngEHT observations provides tentative (though not necessarily conclusive) insight as to whether or not circularity is a principle that could characterize black-hole spacetimes in nature.

In addition, there is a connection of circularity to particular modified-gravity theories: as shown in \cite{Xie:2021bur}, theories that deviate perturbatively from GR give rise to circular spacetimes (under some assumption). Thus, an observation which is well-explained by a non-circular spacetime, hints whether such theories are the correct description of gravity. We add that such hints must be taken with a grain of salt, because they rely on specific assumptions: for instance, the analysis in \cite{Xie:2021bur} made specific assumptions about the branch of solutions of the theory, which may or may not hold in a given theory.

Nevertheless, the connection between specific image features and non-circularity, motivated by a fundamental-physics principle, namely the locality-principle, could be one of several promising gateways to learn about the breakdown of the Kerr paradigm and gain insight into a better description of gravity beyond GR. This motivates us to review parameterizations of black-hole metrics beyond GR, which exist in the literature \cite{1979GReGr..10...79B,Johannsen:2011dh,Johannsen:2013szh,Cardoso:2014rha,Konoplya:2016jvv}, and could be used to search for the breakdown of the Kerr paradigm with EHT or ngEHT observations. Further, it motivates us to propose a more general parameterization, which includes non-circular black-hole spacetimes.

\section{Circular parameterizations}
\label{sec:circular}
We now review that the existing parameterizations in \cite{1979GReGr..10...79B,Johannsen:2013szh, Konoplya:2016jvv} only cover circular spacetimes.
The bumpy Kerr metrics in \cite[Eq.~(2.36)]{Vigeland:2009pr} and \cite[Eq.~(1.5)]{Vigeland:2010xe} contain some non-circular spacetimes but are not exhaustive, in that they only introduce two and three free functions, respectively.
The parameterization in \cite{Johannsen:2013szh} (as well as the implicitly defined spacetimes in \cite[Eqs.~(30)~and~(56)]{Vigeland:2011ji}) 
are circular and additionally assume the existence of a hidden constant of motion.

The construction of the general metric that follows from circularity \cite{1917AnP...359..117W, 1932RSPSA.136..176L, Papapetrou:1966zz,Kundt:1966zz}, also cf.~\cite[Sec.~7.1]{Wald:1984rg}, proceeds as follows. 
To start out, a general stationary and axisymmetric metric has ten non-vanishing and independent metric components. These are functions of the two non-Killing coordinates only, given that we choose the Killing symmetries to be manifest. Because there is the freedom to perform four coordinate transformations, one can always reduce the number of non-vanishing metric components to six in \emph{some} coordinate system. Below, we review how the symmetries of a circular, axisymmetric and stationary spacetime result in five non-vanishing metric components in one \emph{particular} coordinate system, and that those are all but one off-diagonal component.

To be explicit, we work in Boyer-Lindquist coordinates $t, r, \theta, \phi_\text{BL}$, in which $\xi_1= \partial_t$ and $\xi_2=\partial_{\phi_\text{BL}}$, whereby we refer to coordinates which reduce to Boyer-Lindquist coordinates in the limit of Kerr spacetime \footnote{If large deviations from the Kerr spacetime are permitted, such a coordinate choice is no longer unique, because Kerr spacetime can occur more than once in the thus-defined configuration space. This is closely related to what is known as the Gribov problem in non-Abelian gauge theories and it also affects the gravitational configuration space. We therefore restrict ourselves to deviations from Kerr which are small enough such that the Kerr limit can only be taken in one unique way.}.

In the case of a circular spacetime, one has additional isometries, by which metric components can be set to zero, or become functions of one another \cite{Papapetrou:1966zz,Kundt:1966zz, Wald:1984rg}.

For every axisymmetric and stationary spacetime there are ``surfaces of transitivity", labelled by constant values of $r$ and $\theta$, which are generated by $\xi_1$ and $\xi_2$, i.e., the two Killing vectors are tangent to the surfaces. For every circular spacetime, there is a family of 2-surfaces which are everywhere orthogonal to the surfaces of transitivity. Locally, such 2-surfaces exist also if the spacetime is non-circular. Circularity implies that these 2-surfaces, called the meridional surfaces, exist globally \cite{Papapetrou:1966zz, Kundt:1966zz}. There is therefore the additional isometry that is the simultaneous transformation $t \rightarrow -t$ and $\phi_\text{BL} \rightarrow -\phi_\text{BL}$.
Thus, circularity implies that four metric components vanish, namely
\be
0=g_{tr}= g_{t\theta}= g_{\phi_{\rm BL} r}= g_{\phi_{\rm BL} \theta}.\label{eq:circularity}
\ee
Thus, within Boyer-Lindquist coordinates, the most general circular, axisymmetric and stationary spacetime cannot have more than six non-vanishing metric components. However, two additional reductions in the free functions that determine these metric components are possible.
To see this, we focus on the meridional surfaces, which are labelled by  constant values of $t$ and $\phi_{\rm BL}$ and spanned by $r$ and $\theta$. Because of circularity, these surfaces exist globally and we focus on the two-dimensional line element on these: given that every two-dimensional metric is conformally flat, one may always transform from Boyer-Lindquist coordinates $(r,\theta)$ to coordinates $(\tilde{r},\tilde{\theta})$, in which the two-dimensional line element within the meridional surfaces can be written as $ds_{\rm mer}^2 = g_{\tilde{r}\tilde{r}}\left(d\tilde{r}^2 + \tilde{r}^2 d\tilde{\theta}^2 \right)$, with just one free function multiplying the flat 2-d line element in 2-d spherical coordinates. Thus, $g_{\tilde{r}\tilde{\theta}}=0$, reducing the number of non-vanishing metric components to five. Additionally, $g_{\tilde{\theta}\tilde{\theta}}$ is parameterized by the same free function as $g_{\tilde{r}\tilde{r}}$. Thus, there are five non-vanishing metric components, parameterized by four free functions. This is also known as the Lewis-Papapetrou \cite{Papapetrou:1966zz,Kundt:1966zz,Wald:1984rg} form of a circular metric.

As an example, we write the Kerr solution in Lewis-Papapetrou form in App.~\ref{app:Kerr-in-LP-form}.
While this is possible, the resulting coordinates are unconventional. This exemplifies that insisting on the most reduced form may result in coordinates which are unnecessarily complicated.
Because of this reason, it can be advantageous to work in a parameterization in Boyer-Lindquist coordinates, where $g_{rr}$ and $g_{\theta \theta}$ are not the same function.

Such a general spinning black-hole metric (i.e., axisymmetric, stationary and asymptotically flat metric) that respects circularity has been proposed in \cite{Konoplya:2016jvv}.
It takes the form
\bea
ds^2_{RZ}&=& - \frac{f(r, \theta)- \omega(r, \theta)^2 \sin^2\theta}{\kappa^2(r, \theta)} dt^2 - 2 \omega(r, \theta)r \sin^2\theta dt d\phi_\text{BL} + \kappa^2(r, \theta)r^2 \sin^2 \theta d\phi_\text{BL}^2 \nonumber\\
&{}&+ \sigma(r,\theta)\left(\frac{\beta^2(r, \theta)}{f(r, \theta)}dr^2 + r^2 d\theta^2 \right),\label{eq:RZpara}
\eea
with five free functions $f(r, \theta), \beta(r, \theta), \sigma(r, \theta), \kappa(r, \theta)$ and $\omega(r, \theta)$.
The form in Eq.~\eqref{eq:RZpara} makes only partial use of the above coordinate freedom in the 2-surfaces of transitivity by setting $g_{r\theta}=0$. We confirm that one of the free functions in Eq.~\eqref{eq:RZpara} could be removed by a coordinate transformation by explicitly checking that
the circularity conditions Eq.~\eqref{eq:circular1} and \eqref{eq:circular2} hold for any choice of the five functions. \\

Stationary, axisymmetric and circular spacetimes can also contain an additional hidden constant of motion.
Hidden constants of motion are associated to Killing tensors -- the higher-rank generalization of a Killing vector: the defining property of a Killing vector $\xi_\mu$, namely  $\nabla_{(\mu}\xi_{\nu)}=0$, generalizes to a rank-$n$ Killing tensor $K_{\mu_1\dots\mu_n}$, i.e.,
\begin{align}
	\nabla_{(\mu}K_{\mu_1\dots\mu_n)}=0\;,\label{eq:Killingtensor}
\end{align}
where round brackets denote complete symmetrization. 

Clearly, a Killing vector is simply a rank-1 Killing tensor. Moreover, metric compatibility of the covariant derivative based on the Christoffel connection implies that the metric itself is a rank-2 Killing tensor. However, this is not associated to hidden constants of motion, but to the absence of non-metricity degrees of freedom. 
While Killing vectors encode an explicit isometry of the underlying spacetime, more general Killing tensors are only manifest in the local dynamics of test particles, i.e., they imply a hidden constant of motion which can (given sufficiently many other constants of motion) lead to separability of the geodesic equation. The existence of a hidden constant of motion along a geodesic, parameterized by the proper time $\tau$ and with the tangent vector $u^{\mu} =  \frac{d x^\mu(\tau)}{d\tau}$, follows from Eq.~\eqref{eq:Killingtensor} which implies that  the quantity $J =K_{\mu_1....\mu_n} u^{\mu_1}....u^{\mu_n}$ is conserved along a geodesic, i.e., $\frac{d J}{d\tau} = 0$.

The most general metric with two independent Killing vectors and one non-trivial rank-2 Killing tensor
takes the form~\cite{1979GReGr..10...79B}
\begin{align}
\label{eq:Benenti-Francaviglia-ansatz}
	g^{\mu\nu}\partial_\mu \partial_\nu  =\frac{1}{S_{x_1}+S_{x_2}}
	\Big[
		\left(G_{x_1}^{ij}+G_{x_2}^{ij}\right) \partial x_i\partial x_j
		+\Delta_{x_1} \partial x_1^2
		+\Delta_{x_2} \partial x_2^2
	\Big]\;.
\end{align}
Here, the indices $i,\,j$ are associated with the two Killing coordinates. 
The functions $S_{x_1}(x_1)$, $G_{x_1}^{ij}(x_1)$, and $\Delta_{x_1}(x_1)$ as well as $S_{x_2}(x_2)$, $G_{x_2}^{ij}(x_2)$, and $\Delta_{x_2}(x_2)$ depend only on the coordinates $x_1$ and $x_2$, respectively. (We write the inverse metric, since this is the simpler expression.) The resulting non-trivial Killing tensor $K^{\mu\nu}$ and generalized Carter constant $C$ read~\cite{1979GReGr..10...79B}
\begin{align}
	K^{\mu\nu}\partial_{\mu}\partial_{\nu} &= \frac{1}{S_{x_1}+S_{x_2}}
	\Big[
		\left(S_{x_1}G_{x_2}^{ij} - S_{x_2}G_{x_1}^{ij}\right)\partial x_i\partial x_j
		-S_{x_2}\Delta_{x_1}\partial x_1^2
		+S_{x_1}\Delta_{x_2}\partial x_2^2
	\Big]\;,
	\\
	C &= K^{\mu\nu}u_\mu u_\nu\;,
\end{align}
with $u_\mu$ the 4-velocity of the test particle.

The parameterization in Eq.~\eqref{eq:Benenti-Francaviglia-ansatz} is fully equivalent to the one presented in \cite[Eq.~(10)]{Johannsen:2013szh}. We provide the explicit relations in App.~\ref{app:Johannsen-to-BF}. 
We have not  verified a relation to the two parameterizations defined in \cite[Eqs.~(30)~and~(56)]{Vigeland:2011ji}. The difficulty comes about, because the definition of these parameterizations is not explicit but rather given in terms of differential equations. Still, the latter are also built from demanding a rank-2 Killing tensor and \cite{1979GReGr..10...79B} claims generality for this case.
Finally, the Kerr metric (in Boyer-Lindquist coordinates) is also of the above form. In this special case, the hidden constant of motion 
is the Carter constant.

While \cite{1979GReGr..10...79B} does not assume circularity, we have explicitly confirmed that all such metrics are circular. Hence, due to their additional hidden constant of motion, they form a subclass of the most general circular parameterization in Eq.~\eqref{eq:RZpara} written in the form in \cite{Papapetrou:1966zz,Kundt:1966zz,Wald:1984rg}. 

Eq.~\eqref{eq:Benenti-Francaviglia-ansatz} captures the most general stationary and axisymmetric 
metric which exhibits a non-trivial \emph{rank-2} Killing tensor and is, in consequence, circular. 
 Hence, any black-hole  spacetime which cannot be brought into the above form by a suitable coordinate transformation, does not exhibit the associated particular type of hidden constant of motion. At the same time, we are not aware of a proof that precludes the existence of Killing tensors of even higher-rank, cf.~\cite{Owen:2021eez} for an example of a systematic order-by-order (in small spin-parameter and small beyond-GR coupling) search for Killing tensors up to rank 6. It is, to the best of our knowledge, not excluded that axisymmetric, stationary spacetimes with higher-rank (i.e., rank $>2$) Killing tensors are non-circular.

\section{Horizon-penetrating parameterization}
\label{sec:hp}
Horizon-penetrating coordinates can be used to set up black-hole parameterizations in and beyond circularity. There is a strong reason to favor horizon-penetrating coordinates such as ingoing Kerr coordinates $u, r, \chi, \phi$, where $u$ is a lightcone time.
These coordinates make it easy to avoid accidental introductions of curvature singularities at the horizon.  The reason is that Kerr spacetime in horizon-penetrating coordinates does not feature coordinate singularities at the horizon. Therefore, a spacetime that deviates from the Kerr spacetime does not feature curvature singularities at the horizon, as long as the functions that encode these deviations are non-singular and invertible.
In contrast, Boyer-Lindquist coordinates require additional non-trivial conditions on the metric coefficients to achieve the same, cf.~\cite{Johannsen:2013szh, Johannsen:2013rqa, Cardoso:2014rha, Held:2021vwd}. 

To encode deviations from the Kerr spacetime and parameterize more general spacetimes, we write
\bea
ds^2_{\rm HP} &=& - \left(\frac{r^2 -2 M r + a^2 \chi^2}{r^2+a^2\chi^2}\right)\left(1+\Delta_{\rm HP,\,1}(r, \chi) \right) du^2 + 2\left(1+\Delta_{\rm HP,\, 2}(r,\chi) \right) du dr \nonumber\\
&{}&- 4\frac{M a r}{r^2+a^2\chi^2}(1-\chi^2)(1+\Delta_{\rm HP,\, 3}(r, \chi))du d\phi - 2 a (1-\chi^2)(1+\Delta_{\rm HP,\, 4}(r, \chi))dr d\phi\nonumber\\
&{}& + \frac{r^2+a^2\chi^2}{1-\chi^2} (1+\Delta_{\rm HP,\, 5}(r, \chi))d\chi^2 \nonumber\\
&{}&+ \frac{1-\chi^2}{r^2+a^2\chi^2} \left(\left(a^2+r^2 \right)^2- a^2\left(r^2-2M r +a^2 \right)(1-\chi^2) \right) (1+\Delta_{\rm HP,\, 6}(r, \chi))d\phi^2.\label{eq:dsHP}
\eea
In contrast to the circular parameterizations  we have discussed in Sec.~\ref{sec:circular}, we write non-circular spacetimes in terms of deviations from the Kerr spacetime. The reason is phenomenological: there is currently -- within the observational uncertainties -- no indication for deviations of black holes from the Kerr solution, thus deviations are already constrained, cf.~,e.g.,\cite{Held:2019xde,EventHorizonTelescope:2021dqv} for constraints in the context of shadow images. Therefore, writing a parameterization in terms of deviations from Kerr spacetime connects most directly to observations.

We require that the spacetime is asymptotically flat. For $\Delta_{{\rm HP},\, i}=0$, i.e., in the Kerr-limit, this is the case. To preserve this property, we demand that
\be
\Delta_{{\rm HP},\, i}(r, \chi) \overset{r \rightarrow \infty}{\longrightarrow} 0.
\ee
Additionally, one may require that the $\mathcal{O}\left(\frac{1}{r}\right)$ terms agree with those of the Kerr spacetime, such that the Newtonian limit is preserved. To achieve this, the corrections arising from $\Delta_{{\rm HP},\, i}$ must only set in at higher order, i.e., $\Delta_{{\rm HP}, \, i} \sim \mathcal{O}\left(\frac{1}{r^2}\right)$. Similarly, if agreement with the post-Minkowski expansion to higher orders is to be achieved, constraints may be pushed to higher orders.

Next, we consider the limit of flat Minkowski spacetime. For Kerr spacetime, this limit is reached for $M \rightarrow 0$, which results in a Riemann tensor that is identically zero in all its components. For the metric Eq.~\eqref{eq:dsHP}, this is no longer the case. For instance, it suffices to set $\Delta_{{\rm HP},\,2}(r, \chi) \neq 0$, with all other $\Delta_{{\rm HP},\, i\neq2} =0$, for the spacetime to no longer be Ricci flat and feature a non-vanishing Ricci scalar.
To preserve the property that the spacetime is flat in the limit $M \rightarrow 0$, one may demand that $\Delta_{{\rm HP},\, i} \sim M$. Alternatively, the compact object described by Eq.~\eqref{eq:dsHP} may be characterized by additional (quantum) "hair", such that even in the limit $M \rightarrow 0$, a non-trivial spacetime geometry exists. We leave the resulting question, whether or not the parameter $M$ preserves its interpretation as the ADM mass of the compact object for future work.

Similarly, we consider the limit $a \rightarrow 0$, which results in spherical symmetry in the case of Kerr spacetime. This is not the case for Eq.~\eqref{eq:dsHP}, which reduces to 
\bea
ds^2_{\rm HP} &\overset{a \rightarrow 0}{\longrightarrow}& - \left(\frac{r -2 M}{r}\right)\left(1+\Delta_{\rm HP,\,1}(r, \chi) \right) du^2 + 2\left(1+\Delta_{\rm HP,\, 2}(r,\chi) \right) du dr \nonumber\\
&{}& + \frac{r^2}{1-\chi^2} (1+\Delta_{\rm HP,\, 5}(r, \chi))d\chi^2 + \left(1-\chi^2\right)r^2 (1+\Delta_{\rm HP,\, 6}(r, \chi))d\phi^2.\label{eq:dsHPzerospin}
\eea
The remaining $\chi$-dependence in $g_{uu}$ and $g_{ur}$ 
is a clear sign of the breaking of spherical symmetry, as is the deviation of the angular line-element from its canonical form $ds^2_{\rm angular} = \frac{r^2}{1- \chi^2} d\chi^2 + (1- \chi^2)r^2 d\phi^2$. One may object that four coordinate transformations can absorb the additional $\chi$-dependence introduced by the four functions $\Delta_{{\rm HP},\,1/2/5/6}$. However, these coordinate transformations in general cannot be done without new introducing off-diagonal terms in the line element. This can be seen, e.g., by inspecting the curvature invariants of Eq.~\eqref{eq:dsHPzerospin}. As an example, the Ricci scalar is non-vanishing and depends on $\chi$ and $r$ explicitly, as well as through derivatives of $\Delta_{{\rm HP},\,1/2/5/6}$. Therefore, curvature invariants in this limit are  in general not spherically symmetric. 
Accordingly, there are two sources of breaking of spherical symmetry to axisymmetry: one is the presence of spin, $a$, the other is, broadly speaking, additional (quantum) ``hair". This ``hair" is encoded in the $\chi$-dependence of $\Delta_{{\rm HP},\,1/2/5/6}$.
\\

For arbitrary deviations $\Delta_{{\rm HP},\, i}$, changes of the spacetime signature can occur. 
The metric determinant is given by
\bea
{\rm det}(g_{\rm HP})&=& \frac{1+\Delta_5}{1-\chi^2} \Biggl[ \left(1+\Delta_4 \right)a^2\left(1-\chi^2 \right)^2 \Biggl( 2(1+\Delta_2)(1+\Delta_3)M\, r\nonumber\\
&{}&+ (1+\Delta_1)(1+\Delta_4)\left(-2M \,r+r^2+ a^2\chi^2 \right)\Biggr)\nonumber\\
&{}&+ (1+\Delta_2)(1-\chi^2)\Biggl(2(1+\Delta_3)(1+\Delta_4)M a^2 r(1-\chi^2) \nonumber\\
&{}&-(1+\Delta_2)(1+\Delta_6)\left(r^4+\chi^2\, a^2(1+r^2)+2M\, r\, a^2(1-\chi^2) \right)\Biggr)
\Biggr],
\eea
such that the signature translates into conditions on the $\Delta_{{\rm HP},\, i}$.
If all $\Delta_{{\rm HP}\, i} \sim \epsilon $,
this expression simplifies to ${\rm det}(g_{\rm HP}) =-(1+\epsilon)^4(r^2+a^2\chi^2)^2$, which means that the signature does not change as long as $\epsilon>-1$. In fact, the eigenvalues of the metric change their sign at $\epsilon=-1$, such that the metric signature flips from $(-,+,+,+)$ to $(+,-,-,-)$.
\\

The parameterization in Eq.~\eqref{eq:dsHP} reduces to a parameterization of circular black holes, if conditions on the $\Delta_{{\rm HP}, \, i}(r, \chi)$ hold. The circularity conditions in Eq.~\eqref{eq:circular1} and~\eqref{eq:circular2} amount to lengthy differential conditions which are not straightforward to solve.
The only two conditions that we found which are straightforward to solve are: i) $\Delta_{{\rm HP}, \, 5}$ can deviate from zero while preserving circularity, if all other $\Delta_{{\rm HP}, \,i \neq 5}=0$; ii) $\Delta_{{\rm HP},\,i}=\epsilon\,\,\, \forall\,i$ preserves circularity but as soon as $\Delta_{{\rm HP},\,1}$ is chosen to differ from the other deviation functions, circularity is broken.
Accordingly, an explicit restriction to circular spacetimes appears to be quite non-trivial in horizon-penetrating coordinates.
 
Instead, the parameterization in Eq.~\eqref{eq:RZpara} appears to be the preferred one for circular spacetimes, because circularity is straightforward to implement in Boyer-Lindquist coordinates. Thus, it is interesting to understand how the more general parameterization in Eq.~\eqref{eq:dsHP} and the circular parameterization in Eq.~\eqref{eq:RZpara} are related.
An explicit transformation into the parameterization of circular black-hole spacetimes in Boyer-Lindquist coordinates is challenging to provide. Instead, we use a counting argument to plausibilize the existence of such a coordinate transformation. The counting argument adds all available free functions and subtracts the non-trivial conditions that must be satisfied either for the metric to be of the form Eq.~\eqref{eq:RZpara} or for a coordinate transformation to exist.

There are 14 free functions, out of which 6 are the $\Delta_{{\rm HP},\, i}(r, \chi)$ coming from parameterization Eq.~\eqref{eq:dsHP} and 8 are the free functions resulting from coordinate transformations which preserve manifest Killing coordinates. These 14 free functions are subject to 9 constraints, namely 4 differential constraints on the 8 free functions from coordinate transformations, and 5 constraints arising from the vanishing of metric components in the circular metric parameterization in Eq.~\eqref{eq:RZpara}.

That four coordinate transformations, which preserve manifest Killing coordinates, provide 8 free functions, subject to 4 differential constraints, can be seen as follows: A general coordinate transformation from coordinates $x^{\mu}$ to coordinates $y^{\mu}$ can be written as 
\be
dx^{\mu} = F^{\mu}_{\nu}dx^{\nu},
\ee
where the 16 functions $F^{\mu}_{\nu}$ need to form an exact differential and hence are subject to the differential constraints $\partial_{\alpha}F^{\mu}_{\nu} = \partial_{\nu}F^{\mu}_{\alpha}$. Thus, there are 16 free functions subject to 6 differential constraints in this general case. In order to preserve manifest Killing coordinates, the transformations of $r$ and $\chi$ must not involve $u$ and $\phi_{\rm BL}$. If $r$ and $\chi$ were functions of $t$ and $\phi_{\rm BL}$, the metric components would depend on the two Killing coordinates, which would therefore no longer be manifest Killing coordinates.
Further, it must hold that $du = dt+\dots$ and $d\phi= d\phi_{\rm BL} +\dots$, such that the transformation of $u$ and $\phi$ each only contains 2 free functions. Together, this reduces the 16 free functions from the general coordinate transformation to 8 free functions, and the number of differential constraints from 6 to 4, cf.~\cite[Eq.~(2.2)]{1993PhRvD..48.2635G}. 

In addition, 6 free functions are given in the initial form of the metric and 5 constraints arise from the final form of the metric, namely the fact that all but one off-diagonal metric element in Eq.~\eqref{eq:RZpara} vanishes.
 
As a result of 14 free functions with 9 constraints, (at least) 5 free functions remain (more if not all constraints are linearly independent). These functions depend on $r$ and $\chi$.
5 free functions of $r$ and $\chi$ are exactly what is needed to parameterize circular black holes in Boyer-Lindquist coordinates in the form Eq.~\eqref{eq:RZpara}, and are even one function too many if one does not insist on Boyer-Lindquist coordinates, cf.~Sec.~\ref{sec:circular}.

\subsection{Examples of spinning regular black holes included in the horizon-penetrating parameterization}

As two examples of black-hole spacetimes included in Eq.~\eqref{eq:dsHP}, we consider two simpler families of black-hole spacetimes, one of them non-circular, the other circular, but both of them regular.
To obtain a regular black hole, it is sufficient to encode the effect of weakening of gravity. In turn, this is encoded in upgrading the mass parameter $M$ to a spacetime-dependent function. The translation between the general form Eq.~\eqref{eq:dsHP} and the form in 
Eq.~\eqref{eq:dsregloc}, which is non-circular, is given by
\bea
\Delta_{\rm HP,\, 2}&=&0,\label{eq:DeltaHP2}\\
\Delta_{\rm HP,\, 4}&=&0,\\
\Delta_{\rm HP,\, 5}&=& 0,\\
\Delta_{\rm HP,\,1}&=& \frac{2r \left(M-M(r,\chi) \right)}{r^2+a^2\chi^2 - 2M r},\label{eq:Delta1HP}\\
\Delta_{\rm HP, \, 3}&=&\frac{M(r, \chi)- M}{M},\\
\Delta_{\rm HP, \, 6}&=&-\frac{2a^2\left(M(r, \chi)-M \right)r(\chi^2-1)}{r^4+a^4\chi^2+a^2 r \left(2M +r(r-2M)\chi^2 \right)}.\label{eq:localityHP}
\eea
The regular, rotating black holes in, e.g., \cite{Reuter:2006rg,Abdujabbarov:2016hnw,Torres:2017gix,Kumar:2019ohr,Kumar:2020ltt,He:2020dfo,Simpson:2021dyo} can be brought into the form in Eq.~\eqref{eq:dsHP} for the special case $M(r, \chi) = M(r)$ in Eqs.~\eqref{eq:Delta1HP}-\eqref{eq:localityHP}, followed by a coordinate transformation into Boyer-Lindquist coordinates (where for clarity we denote the azimuthal angle $\phi_{\rm BL}$ in Boyer-Lindquist coordinates) according to
\bea
t&=& u - \int dr \frac{r^2+a^2}{r^2+a^2-2M(r) r},\\
\phi_{\rm BL}&=&\phi- \int dr\frac{a}{r^2+a^2-2M(r) r}.
\eea
The fact that the spacetime described by Eqs.~\eqref{eq:DeltaHP2}-\eqref{eq:localityHP} with $M(r, \chi) = M(r)$ is a very special choice in the general class Eq.~\eqref{eq:dsHP} suggests that the regular black holes that have been discussed in the literature are a special subclass of a more general family of rotating regular black holes. Indeed, the black-hole spacetimes with $M(r,\chi)=M(r)$ all fulfill the circularity condition.
\\

Despite having shown examples of non-circular black-hole spacetimes in the parameterization Eq.~\eqref{eq:dsHP}, we have not provided a general proof that all non-circular black-hole spacetimes can be written in this form. Indeed, such a proof is beyond the scope of this paper. Instead,
we motivate the parameterization Eq.~\eqref{eq:dsHP} by the following observations: it contains the Kerr spacetime in the limit $\Delta_{{\rm HP},\,i} \rightarrow 0$. Further, it includes both circular spacetimes, as well as non-circular spacetimes, in particular those that have recently been motivated from a locality principle for new physics~\cite{Eichhorn:2021etc,Eichhorn:2021iwq}. 

We conjecture that the choice of six free functions in the nonvanishing metric components is sufficient to describe all axisymmetric, stationary and asymptotically flat black-hole spacetimes that can be reached as a deformation of the Kerr spacetime. The argument underlying this conjecture is that four free functions can be removed from a general metric by a choice of coordinates, thus six free functions are enough to fully describe a given spacetime. This assumes that the number of free functions corresponds to the number of nonvanishing metric components in horizon-penetrating coordinates.
An interesting example to test this conjecture are the non-circular black-holes in particular scalar-tensor or vector-tensor theories~\cite{Minamitsuji:2020jvf,Anson:2020trg,BenAchour:2020fgy}. These non-circular black holes have been constructed in Boyer-Lindquist coordinates. In \cite[App.~A]{Anson:2020trg}, such non-circular black-holes are transformed (by a combination of coordinate and disformal transformations) to a specific choice of horizon-penetrating coordinates with 7 non-vanishing metric components. 
It remains an open question whether suitable coordinate transformations can be constructed which cast these non-circular black holes to the horizon-penetrating form in Eq.~\eqref{eq:dsHP}. If such a coordinate transformation cannot be found, Eq.~\eqref{eq:dsHP} can be generalized by adding the appropriate deviation function.

In spite of this open question, we argue that Eqs.~\eqref{eq:DeltaHP2}-\eqref{eq:localityHP} with either $M(r, \chi)$ or $M(r)$ are particularly relevant from a quantum-gravity point of view, e.g., 
\cite{Bonanno:2000ep,Reuter:2006rg,Nicolini:2008aj,Modesto:2010rv,Haggard:2014rza,Ashtekar:2018lag,Platania:2019kyx,Nicolini:2019irw,Contreras:2019cmf,Held:2019xde,Eichhorn:2021etc,Eichhorn:2021iwq}, making three of the initial six deviation parameters $\Delta_{\mathrm{HP}, i}$ irrelevant and thus highlighting that Eq.~\eqref{eq:dsHP} does not require further generalization to encode the effects of several quantum-gravity scenarios. The argument is based on the assumption that quantum gravity must regularize spacetime singularities and thus regularize all curvature invariants. A priori, this could occur in distinct ways. One way which has been explored extensively, is captured by Eqs.~\eqref{eq:DeltaHP2}-\eqref{eq:localityHP} and relies on a fast-enough fall-off of the mass function at small $r$. This effectively encodes a weakening of gravity that one can imagine as an effective repulsive force from quantum gravity. From the point of view of black-hole ``hair", the mass parameter is modified to a function, i.e., the corresponding classical ``hair" is modified.

To strengthen the argument that the parameterization in Eqs.~\eqref{eq:DeltaHP2}-\eqref{eq:localityHP} is singled out from a quantum-gravity point of view, we explore a question that, to the best of our knowledge, has not been answered in the literature before, namely: can an alternative variant of ``quantum hair", promoting the spin parameter $a$ to a function $a(r, \chi)$, also result in singularity resolution for spinning black holes? If the answer is yes, it becomes interesting to test whether the resulting spacetimes are circular, thus already captured by the parameterization Eq.~\eqref{eq:RZpara}. If the answer is no, then the special status of Eqs.~\eqref{eq:DeltaHP2}-\eqref{eq:localityHP} is supported.

\subsection{Can hairy spin regularize a black hole?}
\label{sec:hairy-spin}

A different way to give ``quantum hair" to the black hole -- i.e., let it be characterized by more than two free parameters -- is to upgrade $a$ to a spacetime dependent function $a \rightarrow a(r,\chi)$, similarly to \cite{Eichhorn:2021etc,Eichhorn:2021iwq}. The line element of such a metric is given by
\bea
ds^2 &=& - \left(1- \frac{2Mr}{r^2 + a(r,\chi\,)^2 \chi^2} \right)du^2 + 2\, du\,dr - \frac{4M\, a(r,\chi\,)\, r\,}{r^2 + a(r,\chi\,)^2 \chi^2} (1 - \chi^2)\, du\,d\phi \nonumber\\
&{}& - 2 a\,(1 - \chi^2)\, dr\,d\phi + \frac{r^2 +a(r,\chi\,)^2 \chi^2}{1 - \chi^2} d\chi^2 + \frac{1 - \chi^2}{r^2 + a(r,\chi\,)^2 \chi^2}\Biggl[(a(r,\chi\,)^2 + r^2)^2\nonumber\\
&{}&  - a(r,\chi\,)^2 (r^2 - 2 M\,r + a(r,\chi\,)^2) \cdot (1 - \chi^2)\Biggl]\, d\phi^2.
\label{eq:dsmodifiedspin}
\eea
Choosing different functions $a(r,\chi)$ leads to distinct spacetimes. The circularity conditions in Eqs.~\eqref{eq:circular1}-\eqref{eq:circular2} are violated, unless $a(r,\chi) \rightarrow a(r)$, which is similar to our finding for a modification of the mass function  \cite{Eichhorn:2021etc,Eichhorn:2021iwq}. In Boyer-Lindquist coordinates, the parameterization in Eq.~\eqref{eq:RZpara} can thus not capture such spacetimes; instead, additional metric components have to deviate from zero, see Sec.~\ref{sec:BLgen}.

We first explore modifications that satisfy the locality principle from \cite{Eichhorn:2021iwq}, where the spin function $a$ depends on $r,\, \chi$ through the local curvature scale $K_{\rm GR}$ defined in Eq.~\eqref{eq:localcurvature}.

We require that such a locality-based spin function $a(K_{\rm GR})$ satisfies the correct Newtonian limit to leading order, i.e.,
\be
a(K_{\rm GR}) \overset{r \rightarrow \infty}{\longrightarrow} a_{\infty} = \mathrm{const.}\quad \forall\, \chi.
\label{eq:lim_a_large_r}
\ee
In this limit, the curvature scale $K_{\rm GR} \rightarrow 0$. Thus, we demand $a(K_{\rm GR}) \overset{K_{\rm GR} \rightarrow 0}{\longrightarrow}\rm const$.

To regularize black-hole spacetimes, we demand that all curvature invariants are finite everywhere, in particular at the location of the classical singularity, $r \rightarrow 0, \, \chi \rightarrow 0$. We focus on the first curvature invariant $I_1$, defined in terms of the Weyl tensor $C_{\mu \nu \rho \sigma}$ as
\be
I_1 = C_{\mu \nu \rho \sigma} C^{\mu \nu \rho \sigma}.
\label{eq:I_1}
\ee
We determine whether $I_1$ can be made finite everywhere for a locality-based spin function $a(K_{\rm GR})$. This requirement translates into
\be
\lim_{r \rightarrow 0} \lim_{\chi \rightarrow 0} I_1 \overset{!}{=} \lim_{\chi \rightarrow 0} \lim_{r \rightarrow 0} I_1 \overset{!}{<} \infty,
\label{eq:conditions_limits_I1}
\ee
because away from the classical singularity $r \rightarrow 0, \, \chi \rightarrow 0$, no singularities can exist in any of the invariants as long as $a(K_{\rm GR})$ is itself regular. The first curvature invariant Eq.~\eqref{eq:I_1} can be split into its ``classical" part $I_{1, c}$ -- containing no derivatives of $a(K_{\rm GR})$ -- and its counterpart $I_{1, d}$ -- which depends on the derivates of $a(K_{\rm GR})$ -- so that $I_1 = I_{1, c} + I_{1, d}$. The ``classical" part is given by:
\be
I_{1, c} = \frac{48 M^2 \left(r^6 - 15\, a^2(K_{\rm GR})\, r^4 \chi^2 + 15\, a^4(K_{\rm GR})\, r^2 \chi^4 - a^6(K_{\rm GR})\, \chi^6 \right)}{\left(r^2 + a^2(K_{\rm GR})\, \chi^2 \right)^6}.
\label{eq:I_1c}
\ee

We assume that the regularity conditions Eq.~\eqref{eq:conditions_limits_I1} have to apply to $I_{1, c}$, $I_{1, d}$ separately, otherwise delicate cancellations of divergences would have to occur between $I_{1, c}$ and $I_{1,d}$. To derive a necessary condition for singularity resolution, we restrict ourselves to the ``classical'' part $I_{1, c}$.

In the limit $r \rightarrow 0$, we get
\be
 I_{1, c} \rightarrow -\frac{48 M^2}{a^6(K_{\rm GR} \big\vert_{r=0}) \chi^6}. 
 \label{eq:r_lim_I1c}
 \ee
The limit $\chi \rightarrow 0$ of Eq.~\eqref{eq:r_lim_I1c} is finite if the leading behavior of $a \left(K_{\rm GR}(r = 0, \chi \rightarrow 0)\right)$ is of the form $a\left(K_{\rm GR}(r=0, \chi \rightarrow 0)\right) \sim \frac{1}{\chi^n},\, n \geq 1$.

This behavior is indeed fulfilled, if $a(K_{\rm GR}) \sim K_{\rm GR}^p$, $p\geq 1$. This leaves us with the other sequence of limits to check.
Thus, we first set $\chi = 0$. Then, we assume that $a(K_{\rm GR}) \sim K_{\rm GR}^p$, $p\geq 1$, as derived above. In this case
\be
I_{1, c} \rightarrow \frac{48 M^2}{r^6},
\label{eq:chi_lim_I1c}
\ee
which diverges in the limit $r \rightarrow 0$.

Hence, we conclude that  under the assumption specified above, ``quantum hair" $a(K_{\rm GR})$ that depends on $r$ and $\chi$ through the local curvature scale $K_{\rm GR}$ cannot lead to a resolution of the classical singularity. This is an intriguing result, because it implies a certain degree of uniqueness to ``quantum hair" that consists in a mass function  $M(K_{\rm GR})$. Of course, modifications $M(K_{\rm GR})$ and $a(K_{\rm GR})$ can both be present simultaneously in regular black holes, but the modification $a(K_{\rm GR})$ is insufficient on its own, if the locality principle is to simultaneously be satisfied.

This suggests to abandon the locality principle to find regular curvature invariants and look for spin functions $a(r, \chi)$ depending independently on $r$ and $\chi$. We assume that $a(r, \chi)$ has a series expansion starting with $a(r, \chi) \sim r^{\alpha} \chi^{\beta}$, with $\alpha, \beta \in \mathbb{Z}$ for small $r$ and $\chi$. The choice
\be
a(r, \chi) = \chi^{\beta}, \quad \beta \leq -3,
\label{eq:leading_behaviour_spin}
\ee
renders the full curvature invariant $I_1$ finite and single-valued in the limit $r, \chi \rightarrow 0$. This result relies on the absence of subleading terms in $r$.

However, this choice of function $a(r, \chi)$ cannot be a valid spin function since it does not fulfil the condition Eq.~\eqref{eq:lim_a_large_r} that requires $a(r, \chi)$ to become constant at large $r$. Therefore, $r$-dependence needs to be added to Eq.~\eqref{eq:leading_behaviour_spin}. Including an r-dependence in $a(r, \chi)$, while satisfying Eq.~\eqref{eq:leading_behaviour_spin} to leading order in the expansion, one can, for instance, consider rational functions like
\be
a(r,\chi) = a_{\infty} \frac{r}{r + 1} + \frac{1}{\chi^4 (r^3 + 1)},
\label{eq:example_working_spin}
\ee
with $a_{\infty}$ a constant.

Indeed, the spin given in Eq.~\eqref{eq:example_working_spin} satisfies both conditions Eqs.~\eqref{eq:lim_a_large_r}, \eqref{eq:leading_behaviour_spin}. However, for the Ricci scalar and $I_1$ the sequence of limits $\lim_{r\rightarrow 0} \lim_{\chi \rightarrow 0}$ is either indeterminate or divergent. This is due to the presence of subleading terms in the series expansion around $r = 0$ which still depend on r. Considering instead a spin function with exponential suppression in r, such as, e.g.,
\be
a(r, \chi) = \frac{a_{\infty}}{e^{\frac{1}{r}}} \cdot \frac{r}{r+1} + \frac{1}{\chi^4} \left(1 - \frac{1}{e^{\frac{1}{r}}}\right),
\ee 
one observes that these subleading terms remain in the series expansion, rendering the invariants multi-valued and potentially divergent. Indeed, our tests of various functions (with even stronger suppression at small $r$) beyond those reported here suggest that any subleading dependence on $r$ beyond the choice $a(r, \chi) = \chi^{\beta}$ renders curvature-invariants ill-defined.

In summary, in our scoping investigation, we do not find an example of a spin modification that renders curvature invariants finite. As we have not comprehensively explored the space of functions $a(r, \chi)$ that do not satisfy the locality principle, there may potentially be a choice of spin function that results in singularity resolution. 
However, because the spin modification enters both denominators and numerators of different metric coefficients in Eq.~\eqref{eq:dsmodifiedspin} and curvature invariants, rendering all curvature invariants finite seems unachievable.

\section{Boyer-Lindquist coordinates beyond Boyer-Lindquist form}\label{sec:BLgen}

Boyer-Lindquist coordinates are widely used in parameterizations of black-hole spacetimes, and they are well-suited to implement circularity (though not to prevent curvature singularities at the horizon). Thus, we explore how to describe non-circular spacetimes in Boyer-Lindquist coordinates.

To describe non-circular black holes in  Boyer-Lindquist coordinates, one has to go beyond the Boyer-Lindquist form of the metric, i.e., allow metric functions to be nonzero which vanish for a Kerr black hole. Because circularity implies invariance under the simultaneous mapping $t \rightarrow -t,\, \phi_\text{BL} \rightarrow -\phi_\text{BL}$, breaking circularity can be achieved by allowing $g_{t\chi} \neq 0$, $g_{\chi \phi_\text{BL}}\neq 0$, $g_{tr} \neq 0$ or $g_{r \phi_\text{BL}} \neq 0$. Adding these four functions to the four free functions of circular, axisymmetric, stationary spacetime would provide eight free functions. 

In addition, in non-circular spacetimes, meridional surfaces are only locally orthogonal to the surfaces of transitivity, but no longer guaranteed to be integrable. Thus, the argument (cf.~Sec.~\ref{sec:circular}), that reduces the number of free functions in the meridional sector to just one, no longer holds, and that sector has two additional free functions. Thus, in the most general case, ten free functions would a priori be expected. However, it might be possible to extend the local patch in which $ds^2_{\rm mer}= g_{\tilde{r}\tilde{r}} \left(d\tilde{r}^2+\frac{\tilde{r}^2}{1-\chi^2}\,  d\chi^2 \right)$ holds far enough to cover the entire region of a spacetime that one is interested in.

Because one has the freedom to perform four coordinate transformations, we expect that a metric should never have more than six free functions in \emph{some} coordinate system; and indeed, \cite{Ayon-Beato:2007ntk} states that there is a coordinate choice in which six functions are enough. This choice of coordinates could in general depend on the chosen metric.
However, this argument does not take into account that one might want to work in coordinates in which the Killing symmetries are manifest, which limits available coordinate transformations, or indeed fixes the coordinate system to a Boyer-Lindquist one, as we do here. 
Insisting on coordinates in which the Killing symmetries are manifest (and assuming an extension of the local patch, see above), leads to eight free functions, cf.~\cite{1993PhRvD..48.2635G} in the context of global hyperbolicity and a 3+1 decomposition.

Thus, in Boyer-Lindquist coordinates, such a spacetime is described by the line element
\bea
ds^2 &=& - \left(1- \frac{2Mr}{r^2 + a^2 \chi^2}\right) \left(1+\delta_1(r, \chi) \right)dt^2  + 2 \delta_2(r, \chi)dt\,dr+2 \delta_3(r, \chi)dt\, d\chi \nonumber\\
&{}& - \frac{4M r\, a\,(1-\chi^2)}{r^2 + a^2 \chi^2}(1+ \delta_4 (r, \chi))dt\,d\phi_\text{BL}  +  \frac{r^2 +a^2 \chi^2}{r^2 -2 M r + a^2}\left(1+ \delta_5(r, \chi)\right)dr^2 \nonumber\\
&{}&+2\gamma(r, \chi)dr\, d\chi+ 2  \delta_6(r, \chi) dr\, d\phi_\text{BL} + \frac{r^2+a^2 \chi^2 }{1-\chi^2}(1+\delta_7(r, \chi))d\chi^2+2 \delta_8(r, \chi) d\chi d \phi_\text{BL}\nonumber\\
&{}&
+ \left(r^2 +a^2 + \frac{2Mr\, a^2 (1- \chi^2)}{r^2+a^2\chi^2} \right) (1- \chi^2) (1+\delta_9(r,\chi))d \phi_\text{BL}^2.\label{eq:newparaBL}
\eea
If the choice $ds^2_{\rm mer}= g_{rr} \left(dr^2+\frac{r^2 -2 M r + a^2}{1-\chi^2}\,  d\chi^2 \right)$, which is always possible locally, is available in a large enough patch of spacetime, then $\gamma(r,\chi)=0$ and additionally $\delta_5(r,\chi)$ and $\delta_7(r,\chi)$ are related to each other, leaving a total of 8 free functions.

For $\delta_i(r,\chi)=0$, the metric reduces to the Kerr metric in Boyer-Lindquist coordinates. Increasing values of $\delta_i(r, \chi)$ parameterize deviations from Kerr spacetime. To ensure that the spacetime remains asymptotically flat, we require
\be
\delta_i(r,\chi) \overset{r \rightarrow \infty}{\longrightarrow} 0.
\ee
Regularity imposes differential conditions on the $\delta_i(r, \chi)$, both at $r=0=\chi$ as well as at $r= r_{+}$, the location of the event horizon. Whereas conditions at $r=0$ might have been expected, conditions at finite $r = r_+$ are not immediately obvious. These arise, because for $\delta_i=0$, the metric Eq.~\eqref{eq:newparaBL} contains the well-known coordinate singularities of Kerr spacetime in Boyer-Lindquist coordinates. These singularities in metric components cancel in curvature invariants. This cancellation requires that different metric functions are delicately balanced. Thus, arbitrary deformations of metric functions, i.e., arbitrary choices of $\delta_i$, can easily introduce curvature singularities which lie on the horizon, i.e., naked singularities, cf.~\cite{Johannsen:2013szh, Johannsen:2013rqa, Cardoso:2014rha, Held:2021vwd}.

We now show explicitly that the parameterization Eq.~\eqref{eq:newparaBL} contains all circular black holes and also several examples of non-circular black holes.

The more specialized parameterization from \cite{Konoplya:2016jvv}, see Eq.~\eqref{eq:RZpara}, that respects circularity, is of course included in Eq.~\eqref{eq:newparaBL}. This is easiest to see by switching from $\chi = \cos\theta$ back to $\theta.$ Then, the mapping between Eq.~\eqref{eq:RZpara} and Eq.~\eqref{eq:newparaBL} is given by:
\bea
\omega &=& \frac{2 a M(1+\delta_4)}{r^2+a^2 \cos^2\theta},\nonumber\\
\kappa^2 &=& \frac{1+\delta_9}{r^2} \left(a^2+r^2 + \frac{2M \, r\, a^2 \sin^2\theta}{r^2+a^2 \cos^2\theta} \right),\nonumber\\
\sigma &=& \left(\frac{a^2 \cos^2\theta}{r^2}+1\right)\cdot \left(1+ \delta_7  \right),\nonumber\\
f&=& \frac{1}{r^2\left(r^2+a^2 \cos^2\theta \right)^2} \Biggl[r^3(a^2+r^2)(r-2M(1+\delta_1))(1+\delta_9) + a^2\Biggl[ a^2(a^2+r^2)(1+\delta_9)\cos^4\theta \nonumber\\
&{}&-r(1+\delta_9)\cos^2\theta\left(2r^2(M-r+M \delta_1)+ a^2 (M-2r+2M \delta_1) +a^2 \cos(2\theta) M\right)\nonumber\\
&{}&+ 2M r^2\left(r(1+\delta_9)-2M(\delta_1 -2 \delta_4 - \delta_4^2+\delta_9+\delta_1\delta_9)\right) \sin^2 \theta
\Biggr]\Biggr], \nonumber\\
\beta^2 &=&\frac{1+\delta_5}{(a^2+r(r-2M))(1+\delta_7)(r^2+a^2\cos^2\theta)^2} \cdot \nonumber\\
&{}& \cdot \Biggl[ r^3(r^2+a^2)(r-2M(1+\delta_1))(1+\delta_9)+ a^2\Biggl[ a^2(a^2+r^2)(1+\delta_9)\cos^4\theta \nonumber\\
&{}&- r(1+\delta_9)\cos^2\theta\left(2r^2(M-r+M\delta_1) +a^2(M-2r+2M \delta_1)+a^2 M \cos(2\theta)\right)\nonumber\\
&{}& +2M r^2\left(r(1+\delta_9)-2M(\delta_1-2 \delta_4-\delta_4^2+\delta_9+ \delta_1 \delta_9) \right)\sin^2\theta
\Biggr]
\Biggr],
\eea
with $\delta_{2,3,6,8}=0$.

Next, we consider the spacetimes from \cite{Eichhorn:2021etc,Eichhorn:2021iwq}, see Eq.~\eqref{eq:dsregloc}. These cannot be represented in Boyer-Lindquist form (i.e., with five non-vanishing metric components) as in \cite{Konoplya:2016jvv}, because that spacetime is non-circular, as we have confirmed by an explicit calculation. 
Here, we show that it can be written in Boyer-Lindquist coordinates if $g_{t\chi}\neq0$ and $g_{\chi\phi}\neq 0$ and that it thus is contained in Eq.~\eqref{eq:newparaBL}.
We first write an ansatz for a coordinate transformation 
\bea
du &=&  dt + \mathcal{F}_r\, dr + \mathcal{F}_{\chi}\, d\chi,\\
d\phi &=& d\phi_{\rm BL}+ \mathcal{G}_r dr + G_{\chi}d\chi, \label{eq:trafo2}
\eea
from the horizon-penetrating coordinates $u, r, \chi, \phi$ to Boyer-Lindquist coordinates $t, r, \chi, \phi_{\rm BL}$. We require that 	$g_{r\chi}= 0=g_{r\phi_{\rm BL}}$, i.e., $\delta_6=0$. These conditions can be solved by requiring that
\be
\mathcal{F}_r = \frac{r^2+a^2}{r^2+a^2- 2r\, M[r, \chi]},\quad \mathcal{G}_r = \frac{a}{r^2+a^2 - 2r\, M[r, \chi]}.
\ee
Because $\mathcal{F}_{\chi}$ and $\mathcal{G}_{\chi}$ are unrestricted, they can be chosen such that $\partial_r \mathcal{F}_{\chi} = \partial_{\chi}\mathcal{F}_r$ and $\partial_r \mathcal{G}_{\chi} = \partial_{\chi}\mathcal{G}_r$. This means that the differential forms in Eq.~\eqref{eq:trafo2} are exact. This does not work in the case where we additionally require $g_{t\chi}= 0$, see App.~A.~5 in \cite{Eichhorn:2021iwq}.
In our case, we can write
\bea
t&=& u - \int dr \frac{r^2+a^2}{r^2+a^2- 2M(r,\chi) r},\\
\phi_{\rm BL}&=& \phi - \int dr \frac{a}{r^2+a^2- 2M(r, \chi) r},
\eea
which for the case $M(r, \chi)= m = \rm const$ reduces to the standard transformation between ingoing Kerr and Boyer-Lindquist coordinates, where it results in $g_{t\chi}=0$. In addition, asymptotically, where $M(r \rightarrow \infty, \chi) \rightarrow m$, the coordinates $t,r, \chi, \phi_{\rm BL}$ form the standard coordinate system of asymptotically flat spacetime in Boyer-Lindquist form. The non-circular, locality-principle-based black-hole spacetime in \cite{Eichhorn:2021iwq} can thus be written as 
\bea
ds^2_{\rm reg,\, local}
&=&
	- \Bigg[1- \frac{2M(r,\chi) r}{\Sigma(r,\chi)} \Bigg]dt^2
	+ \Bigg[\frac{\Sigma(r,\chi)}{\overline{\Delta}(r,\chi)}\Bigg]dr^2
	+ \Bigg[\frac{4M(r,\chi)\,r\,a(\chi^2-1)}{\Sigma(r,\chi)}\Bigg] dt d\phi_{\rm BL}
	\nonumber\\
	&{}&
	+ \Bigg[
		-\frac{\Sigma(r,\chi)}{\chi^2-1}
		-\mathcal{M}_2(r,\chi)^2
		+ \mathcal{M}_1(r,\chi)^2\,\left(r^2+a^2 \right)(\chi^2-1)
		+ \frac{2M(r, \chi)r}{\Sigma(r,\chi)} \mathcal{M}(r,\chi)^2
	\Bigg]d\chi^2
	\nonumber\\
	&{}&
	+ \Bigg[
		\left(r^2 + a^2\right)\left(\chi^2 -1\right)
		-\frac{2M(r, \chi)r\,a^2\left(\chi^2 -1\right)^2}{\Sigma(r,\chi)} 
	\Bigg] d\phi_{\rm BL}^2
	\nonumber\\
 	&{}& 
 	+ \Bigg[
 		\frac{4M(r, \chi)\,r\mathcal{M}(r,\chi)}{\Sigma(r,\chi)}
 		-2\mathcal{M}_2(r,\chi) 
	\Bigg] dt d\chi
	\nonumber\\
	&{}&
	+ \Bigg[ 
		\frac{4M(r, \chi)\,r\,a\,\mathcal{M}(r,\chi)(\chi^2-1)}{\Sigma(r,\chi)}
		- 2(\chi^2-1)(r^2 + a^2)\mathcal{M}_1(r,\chi)  
	\Bigg] d\chi d\phi_{\rm BL}.
	\label{eq:localBL}
	\eea
	with
	\bea
\Sigma(r,\chi) &=& (r^2 + a^2\chi^2)\;,
\quad\quad
\mathcal{M}(r,\chi) = \mathcal{M}_2(r,\chi)+\mathcal{M}_1(r,\chi)(\chi^2-1)\,a\;,
\\[1.5em]\notag
\overline{\Delta}(r,\chi) &=& (r^2 -2\,M(r, \chi)\,r +a^2),\\[1.5em]\notag
\mathcal{M}_1(r,\chi) &=& \frac{d}{d\chi}\int dr\,\frac{a}{r^2 -2\,M(r, \chi)\,r +a^2}\;,
\quad\quad
\mathcal{M}_2(r,\chi) = \frac{d}{d\chi}\int dr\,\frac{r^2 + a^2}{r^2 -2\,M(r, \chi)\,r +a^2}\;,
\eea
This form of the metric exemplifies that in order to avoid the generation of curvature singularities at the event horizon, deviations from the Kerr spacetime have to take a somewhat intricate form, because delicate cancellations are necessary. 
In comparison, the metric in horizon-penetrating coordinates, cf.~Eq.~\eqref{eq:dsregloc}, takes a much simpler form, therefore also enabling faster and more efficient manipulation (e.g., for the calculation of curvature invariants).

\section{Conclusions and outlook}\label{sec:conclusions}
The EHT has an opportunity to probe the Kerr paradigm. To that end, it is important to understand which image features we may look for as indications of the breakdown of GR. Here, one may look to specific theories beyond GR, as well as to a more general principled-parameterized approach, to identify promising image features that may be connected to specific spacetime properties. Such explicit connections may inspire a corresponding search for the respective image features in the observational data, assuming that the overall instrument sensitivity and resolution is sufficient. 

Here, we discover that the black-hole spacetimes introduced in \cite{Eichhorn:2021iwq} are not circular. This gives rise to a (not necessarily one-to-one) correspondence between the specific deviations from circularity and the image features discovered in \cite{Eichhorn:2021etc,Eichhorn:2021iwq}, cf.~Sec.~\ref{sec:cuspsanddents}, which consist of 
cusps, a dent and an asymmetry in the photon rings surrounding the black-hole shadow.
To the best of our knowledge, circular black-hole spacetimes in the literature exhibit some, but not all of these features in combination.

This insight further motivates us to review parameterizations of axisymmetric, stationary and asymptotically flat black-hole spacetimes. Circular spacetimes, due to the additional isometry that is implied by circularity, parameterize black-hole spacetimes with just four free metric functions that occur in five non-vanishing metric components.
We review how \cite{Johannsen:2013szh,Konoplya:2016jvv} are based on circularity, and therefore do not accommodate non-circular spacetimes, such as  \cite{Vigeland:2009pr,Vigeland:2010xe,Minamitsuji:2020jvf,Anson:2020trg,BenAchour:2020fgy,Eichhorn:2021etc,Eichhorn:2021iwq}. We then generalize to a parameterization of black-hole spacetimes beyond circularity. Our preferred form of this parameterization is provided in horizon-penetrating coordinates, where we write it with six metric functions. 
This choice of coordinates enables us to write deviations from the Kerr metric without introducing spurious curvature singularities at the horizon. In contrast, a similar parameterization in Boyer-Lindquist coordinates must satisfy differential identities in order to ensure the absence of singularities. Further, giving up circularity in Boyer-Lindquist coordinates can in general lead to four additional non-zero metric components.

We discuss some of the properties of the non-circular black-hole spacetimes and also show how the example \cite{Eichhorn:2021etc,Eichhorn:2021iwq} can be transformed into Boyer-Lindquist coordinates, if one goes beyond circular parameterizations and allows additional metric components to be nonzero.

Within the family of non-circular spacetimes, we also explore an alternative to a popular class of modifications: whereas numerous examples exist that show that promoting the mass parameter to a mass function can give rise to a regular spacetime \cite{Bardeen:1968,Dymnikova:1992ux,Bonanno:2000ep,Hayward:2005gi,Simpson:2019mud}, we test whether the same can be achieved by promoting the spin parameter to a spin function. We find indications that this is not the case. Thus, regular black holes which are based on a mass function play a special role.
\\

Several questions for future work follow from our investigations.
A first open question is, what the minimal, general parameterization of axisymmetric, stationary and asymptotically flat black-hole spacetimes is. It is at present unclear, whether the parameterization in terms of ten (or, under additional assumptions, eight) deviation functions that we present in Boyer-Lindquist coordinates, overparameterizes the space of these spacetimes. Similarly, it is unclear whether the parameterization in terms of six deviation functions in horizon-penetrating coordinates is general enough to capture all such spacetimes. A general counting argument suggests that it is, but the counting argument does not consider that a given set of coordinates may not cover the full spacetime, but just a patch of it. 
However, one guiding rationale in the present paper was the necessity to find a parameterization which can account for the black-hole spacetimes in \cite{Eichhorn:2021etc,Eichhorn:2021iwq}. 

A second open question is the relation of the two parameterizations of non-circular spacetimes that we have provided.
There are good arguments for and against the use of Boyer-Lindquist coordinates. The ease with which circularity can be imposed and the numerous previous studies on properties of parameterized black holes in these coordinates are a strong argument in favor. However, the difficulty in avoiding spurious curvature singularities at the horizon, which, in general, results in differential constraints on metric components, is a strong argument against. Therefore, we have provided parameterizations in two different sets of coordinates. 
However, providing a general coordinate transformation relating the two without specifying the deviation functions is not possible. Thus, we do not know how the six deviation functions in horizon-penetrating coordinates transform into the ten (or eight) deviation functions in Boyer-Lindquist coordinates, except in special cases.

A third open question is whether non-circular spacetimes may have hidden constants of motion based on a higher-rank Killing tensor. For a rank-two Killing tensor, circularity need not be assumed, but comes out automatically.  This may be different for rank higher than two. Moreover, the presence of a rank-two Killing tensor ensures the separability of the geodesic equation of a test particle and, hence, its analytical solvability. This might be viewed as a drawback from a calculational point of view, but does not imply that non-circular spacetimes are not phenomenologically relevant -- their phenomenology may just be more challenging to characterize.

Finally, on a phenomenological level, we do not know whether the connection between deviations from circularity and the dent- and cusp-like image features as well as the image asymmetry (cf.~Fig.~\ref{fig:noncircular-vs-circular}) is more general than just for the family of spacetimes in \cite{Eichhorn:2021etc,Eichhorn:2021iwq}. There are examples of circular spacetimes in the literature which exhibit one of these features, but not all three in combination. 
In the future, it will be highly interesting to understand, whether such a combination of features can be constructed in a circular spacetime as well or whether it is indeed a unique imprint of specific deviations from circularity. 

Our work lays the basis for future work to derive image features and constrain deviation parameters of spacetime metrics beyond Kerr, such as the metric given in Eq.~\eqref{eq:dsHP}. First steps in this direction have recently been made, both for circular and non-circular spacetimes. A particular case of the non-circular spacetime in Eq.~\eqref{eq:dsHP}, given in Eq.~\eqref{eq:dsregloc}, has been studied in \cite{Eichhorn:2021etc,Eichhorn:2021iwq}.
Restricting to circular spacetime metrics, which pass post-Newtonian constraints from observations in the solar-system~\cite{Will:2014kxa}, two different approaches have been followed in the literature. In \cite{Ayzenberg:2022twz, Younsi:2021dxe} the authors obtained images of the black-hole shadow and photon rings with current EHT capabilities. They provided models to account for astrophysical uncertainties linked to parameters in the accretion disk, and placed constraints on parameters characterizing circular deviations from Kerr, accounting for astrophysical uncertainties linked to parameters of the accretion disk.
Another approach has been taken in \cite{Shashank:2021giy, Cardenas-Avendano:2019zxd}, where constraints on deviation parameters were derived using gravitational wave data from the LIGO/VIRGO collaboration on the inspiral phase of black-hole mergers. Going forward, both approaches may be useful. However, it has to be kept in mind that constraints obtained from 
binary-black-hole mergers only constrain the metrics of supermassive black holes observed by the EHT if one assumes that a black-hole uniqueness theorem holds. In theories beyond GR, this may not be the case, thus constraints from both types of observations are valuable.

\begin{acknowledgments}   
We thank David McNutt for discussions. This work is supported by a research grant (29405) from VILLUM fonden. The work leading to this publication was supported by the PRIME programme of the German Academic Exchange Service (DAAD) with funds from the German Federal Ministry of Education and Research (BMBF).
During parts of this project, A.~H. was supported by a Royal Society International Newton Fellowship under the grant no. NIF\textbackslash R1\textbackslash 191008. 
 \end{acknowledgments}

\appendix

\section{Equivalence of parameterizations with a hidden constant of motion}
\label{app:Johannsen-to-BF}

Here, we present the explicit relations between the parameterization in Eq.~\eqref{eq:Benenti-Francaviglia-ansatz} and the one presented in \cite[Eq.~(10)]{Johannsen:2013szh}. For convenience, we repeat the form of the latter, i.e.,
\begin{align}
	g^{\alpha\beta} \partial_\alpha \partial_\beta 
	=& 
	-\frac{1}{\Delta\,\tilde{\Sigma}} \Big[
		(r^2+a^2)A_1(r)\partial_t 
		+ a A_2(r) \partial_\phi 
	\Big]^2 
	+ \frac{1}{\tilde{\Sigma}\sin(\theta)^2} \Big[
		A_3(\theta)\partial_\phi 
		+ a \sin(\theta)^2 A_4(\theta)\partial_t 
	\Big]^2 
	\notag\\
	& + \frac{\Delta}{\tilde{\Sigma}} A_5(r)\left( \partial_r \right)^2 
	+ \frac{1}{\tilde{\Sigma}} A_6(\theta)\left( \partial_\theta \right)^2\;,
\end{align}
where $f(r)$, $g(\theta)$, $A_i(r)$, $i=1,2,5$ and $A_j(\theta)$, $j=3,4,6$ are functions of only $r$ or $\theta$, respectively. Moreover, $\Delta = r^2 -2Mr+a^2$ and $\Sigma=r^2 + a^2\cos(\theta)^2$ denote the common functions appearing also in the Kerr metric in Boyer-Lindquist coordinates (with $M$ and $a$ the asymptotic black hole mass and spin, respectively) and $\tilde{\Sigma} = \Sigma + f + g$. 

Identifying $x_i, x_j$ with Killing coordinates $t,\phi$ and $(x_1,x_2)=(r,\theta)$, the relations
\begin{align}
	S_r &= \Sigma + f\;,
	\\
	S_\theta &= g\;,
	\\
	\Delta_r &= \Delta\,A_5\;,
	\\
	\Delta_{\theta} &= A_6\;,
	\\
	G_r^{tt} &= -\frac{(r^2 + a^2)^2 A_1^2}{\Delta}\;,
	\\
	G_r^{t\phi} &= -\frac{(r^2 + a^2)a\,A_1\,A_2}{\Delta}\;,
	\\
	G_r^{\phi\phi} &= -\frac{a^2\,A_2^2}{\Delta}\;,
	\\
	G_\theta^{tt} &= a^2\,\sin(\theta)^2\,A_4^2\;,
	\\
	G_\theta^{t\phi} &= a\,A_3\,A_4\;,
	\\
	G_\theta^{\phi\phi} &= \frac{A_3^2}{\sin(\theta)^2}\;,
\end{align}
establish the equivalence of the two parameterizations.

\section{Kerr spacetime in Lewis-Papapetrou form}
\label{app:Kerr-in-LP-form}

In this appendix, we explicitly show that Kerr spacetime can be written in Lewis-Papapetrou form with coordinates $(t,\tilde{r},\theta,\phi_\text{BL})$. In particular, we demand that, besides the three non-vanishing metric components $g_{tt},\,g_{t\phi}\,$ and $g_{\phi_\text{BL}\phi_\text{BL}}$, the line element on the meridional surfaces takes the form $ds_{\rm mer}^2 = g_{\tilde{r}\tilde{r}}\left(d\tilde{r}^2 + \tilde{r}^2 d \theta^2 \right)$, with just one free function $g_{\tilde{r}\tilde{r}}$. The difference to Boyer-Lindquist coordinates $(t,r,\theta,\phi_\text{BL})$ lies in the condition that relates the $g_{\tilde{r}\tilde{r}}$ and $g_{\theta\theta}$, i.e.,
\begin{align}
	\label{eq:cond-LP}
	g_{\theta\theta} = \tilde{r}^2\,g_{\tilde{r}\tilde{r}}\;.
\end{align}
We start from Kerr spacetime in Boyer-Lindquist coordinates and allow for an unknown coordinate transformation $r(\tilde{r})$ of the radial Boyer-Lindquist coordinate $r$. Under this transformation, the above condition \eqref{eq:cond-LP} provides a differential equation
\begin{align}
	\frac{dr}{d\tilde{r}} = \frac{\Delta(r(\tilde{r}))}{\tilde{r}^2}\;,
\end{align}
with $\Delta(r)=r^2 - 2\,M\,r + a^2$ in Boyer-Lindquist coordinates. There are two solutions to this differential equation,
\begin{align}
	r(\tilde{r}) &= 
	\frac{e^{-c_1}}{2\tilde{r}}\left(
		M^2 
		- a^2 
		+ 2\,e^{c_1}M\,\tilde{r} 
		+ e^{2\,c_1}\tilde{r}^2
	\right)\;,
	\\
	r(\tilde{r}) &= 
	\frac{e^{-c_2}}{2\tilde{r}}\left(
		M^2\,\tilde{r}^2 
		- a^2\,\tilde{r}^2 
		+ 2\,e^{c_2}M\,\tilde{r} 
		+ e^{2\,c_2}
	\right)\;,
\end{align}
with constants of integration $c_1$ and $c_2$, respectively. Since the two solutions are related by $c_2 = \log(-e^{-c_1}(M^2 - a^2))$, we focus only on the first solution in the following. For simplicity, we pick $c_1=0$. For the resulting coordinate transformation we pick one of the two branches, which is invertible outside the event horizon,
\be
\tilde{r}(r) =r - M +\sqrt{r^2 - 2\,M\,r + a^2}\;,
\ee
\begin{align}
\label{eq:LP-to-BL}
		- a^2
	dr = 
	\frac{
		a^2 - M^2 + \tilde{r}^2
	}{2\,\tilde{r}^2}\,d\tilde{r}
	\;.
\end{align}
This relates Kerr spacetime in Boyer-Lindquist coordinates $(t,r,\theta,\phi)$ to Kerr spacetime in coordinates $(t,\tilde{r},\theta,\phi_\text{BL})$ which realize the desired form, i.e.,
\begin{align}
\text{ds}^2 =&
	-\text{dt}^2
	+\frac{4 M \tilde{r} \left((M+\tilde{r})^2-a^2\right)}{\left(a^2-(M+\tilde{r})^2\right)^2+4 a^2 \tilde{r}^2 \cos^2(\theta )}
	\left(\text{dt}-a\sin ^2(\theta)\text{d$\phi_\text{BL}$}\right)^2
	\notag\\&
	+\left(
		\frac{\left(a^2-(M+\tilde{r})^2\right)^2}{4\tilde{r}^2}+a^2
	\right)\sin ^2(\theta )\text{d$\phi_\text{BL}$}^2
	\notag\\&
	+\frac{\left(\left(a^2-(M+\tilde{r})^2\right)^2+4 a^2\tilde{r}^2 \cos ^2(\theta )\right)}{4 \tilde{r}^4}
	\left(\text{d$\tilde{r}$}^2 + \tilde{r}^2\text{d$\theta $}^2\right)\;.
\end{align}

\bibliography{References}
\end{document}